\documentclass[12pt]{iopart}
\usepackage{graphicx}
\usepackage{iopams}
\usepackage{textcomp}
\usepackage{xcolor}

\begin{document}

\title[ESR on NaYbS$_{2}$]{Electron Spin Resonance on the spin-1/2 triangular magnet NaYbS$_{2}$}

\author{J\"org Sichelschmidt$^1$, Philipp Schlender$^2$, Burkhard Schmidt$^1$, M. Baenitz$^1$ and Thomas Doert$^2$}

\address{$^1$ Max Planck Institute for Chemical Physics of Solids, Dresden, Germany}
\address{$^2$ TU Dresden, Dept. of Chemistry and Food Chemistry, Dresden, Germany}
\ead{Sichelschmidt@cpfs.mpg.de}

\begin{abstract}

The delafossite structure of NaYbS$_{2}$ contains a planar spin-1/2 triangular lattice of Yb$^{3+}$ ions and features a possible realisation of a quantum spin-liquid state. We investigated the Yb$^{3+}$ spin dynamics by Electron Spin Resonance (ESR)  in single-crystalline samples of NaYbS$_{2}$. Very clear spectra with a well-resolved and large anisotropy could be observed down to the lowest accessible temperature of 2.7~K. In contrast to the ESR properties of other known spin-liquid candidate systems, the resonance seen in NaYbS$_{2}$ is accessible at low fields ($<1$T) and is narrow enough for accurate characterisation of the relaxation rate as well as the $g$ factor of the Yb$^{3+}$ spins.

\end{abstract}

\section{Introduction}

Frustrated spin systems, in particular spins on a triangular lattice, were early on discussed to harbour a quantum spin-liquid (QSL) state \cite{anderson73a,balents10a}. 
Despite the large amount of experimental and theoretical work on the QSL state \cite{savary17a}, there is still no consensus on its existence because the experimental confirmation of a QSL state still remains controversial. 
Two key features for a QSL state, namely the absence of magnetic order and persisting magnetic fluctuations down to lowest temperatures, have been found for organic triangular-lattice solids \cite{shimizu03a,yamashita08a,yamashita10a} as well as for solid-state Kagome \cite{helton07a} or hyper-Kagome systems \cite{okamoto07a}.

Charge-neutral excited states of a spin liquid lead to a constant magnetic susceptibility and a heat capacity which is linear in temperature. These features, usually associated with metals, are found in various insulating QSL model systems.
Therefore, besides verifying the absence of magnetic order at comparatively low temperatures, studying magnetic excitations directly with spin-sensitive probes is one promising way to identify a QSL state in a material.   
It was shown that Electron Spin Resonance (ESR)  is an appropriate low-energy, local spectroscopic probe for spinons in a QSL with significant spin-orbit coupling \cite{luo18a}. It is expected that cooling towards temperatures which correspond to the exchange energy leads to a broadening of the spinon resonance. 
	
Several QSL candidate materials have been investigated by ESR \cite{zorko08a,kermarrec14a,padmalekha15a,li15a}.
Noteworthy, the Yb$^{3+}$-ESR in the triangular lattice system YbMgGaO$_{4}$ was investigated in great detail \cite{li15a,li15b,li17b}. However, the random Mg/Ga occupancy in the layers around the YbO$_{6}$ octahedra \cite{li17b} leads to broadened crystal field excitations and an associated considerable broadening of the ESR line. Thus, the spectra could not be fully resolved which impeded a reliable extraction of the line parameters \cite{li15a}.

The recently explored NaYbS$_{2}$ is unique for it is hosting an effective spin-1/2 quantum magnet on a \textit{perfect} triangular lattice with no inherent structural distortions and atomic site disorder \cite{baenitz18a}. The planar triangular spin arrangement in the delafossite structure allows for a strongly anisotropic quasi-twodimensional magnetism which shows no long-range magnetic order at least down to 260~mK. We report a clear Yb$^{3+}$ spin resonance which is well-resolved and thus provides important local information of the Yb$^{3+}$ $g$ factor and spin dynamics. 

\section{Experimental}
\subsection{Electron Spin Resonance}
Electron Spin Resonance (ESR) probes the absorbed power $P$ of a transversal magnetic microwave field $b_{mw}$ as a function of an external magnetic field $\mu_0H=B$. To improve the signal-to-noise ratio, a lock-in technique is used by modulating the static field, which yields the derivative of the resonance signal d$P$/d$B$. The ESR experiments were performed at X-band frequencies ($\nu $=9.4 GHz) using a continuous-wave ESR spectrometer. 
The sample temperature was set with a helium-flow cryostat allowing for temperatures between 2.7 and 300~K \cite{campbell76a}.

The obtained spectra were fitted by a Lorentzian line shape yielding the parameters linewidth $\Delta B$ which is a measure of the spin-probe relaxation rate, and resonance field $B_{\rm res}$ which is determined by the effective $g$-factor $(g=h\nu/\mu_{\rm B}B_{\rm res}$) and internal fields. The ESR intensity $I_{ESR}$ is determined by the static spin-probe susceptibility $\chi_{\rm ESR}$ along the microwave magnetic field. Thus, $I_{ESR}$ provides a direct microscopic probe of the sample magnetization. $I_{ESR}$ is related to the integrated ESR absorption $I_{A}$ as $I_{ESR}=I_{A}g$ where $g$ is the g-value component along $B$ \cite{gruner10a}. $I_{A}$ was calculated using the line amplitude and line width as reported earlier \cite{wykhoff07b}. 

\subsection{Sample preparation and characterisation}

NaYbS$_{2}$ crystallises in a delafossite structure with trigonal $R\bar{3}m$ symmetry featuring two layers of triangular lattices. The Na$^{+}$ ions reside in the ``A site'' layer whereas the Yb$^{3+}$ ions are located in YbS$_{6}$ octahedra within the ``B site'' layer. 
Yb$^{3+}$ occupies a single crystallographic site with $\bar{3}m$ symmetry, i.e. a site with inversion symmetry  \cite{schleid93a}, in contrast to YbMgGaO$_{4}$ where it resides on a $3m$ site without inversion symmetry \cite{li15a}. 

The ESR experiments were performed on crystals of NaYbS$_{2}$ (two batches) and on polycrystals (one batch). 
Crystals and polycrystalline samples were obtained via different reaction routes; the latter may contain small amounts of Yb$_{2}$O$_{2}$S as impurity phase \cite{baenitz18a}.  
As shown in Fig. \ref{Fig1} the two crystal batches mostly differ in the size of the crystals: they are transparent green-yellowish thin ($100\,\mu$m) platelets with dimensions  of $2\times1$~mm$^{2}$ (``crystal 1'') and $0.3\times0.3$mm$^{2}$ (``crystal 2''). The larger one shows black thin lines which indicate macroscopic domains due to crystals growing on top of each other. 

\section{Results}

\begin{figure}[h]
\begin{center}
\includegraphics[width=0.8\columnwidth]{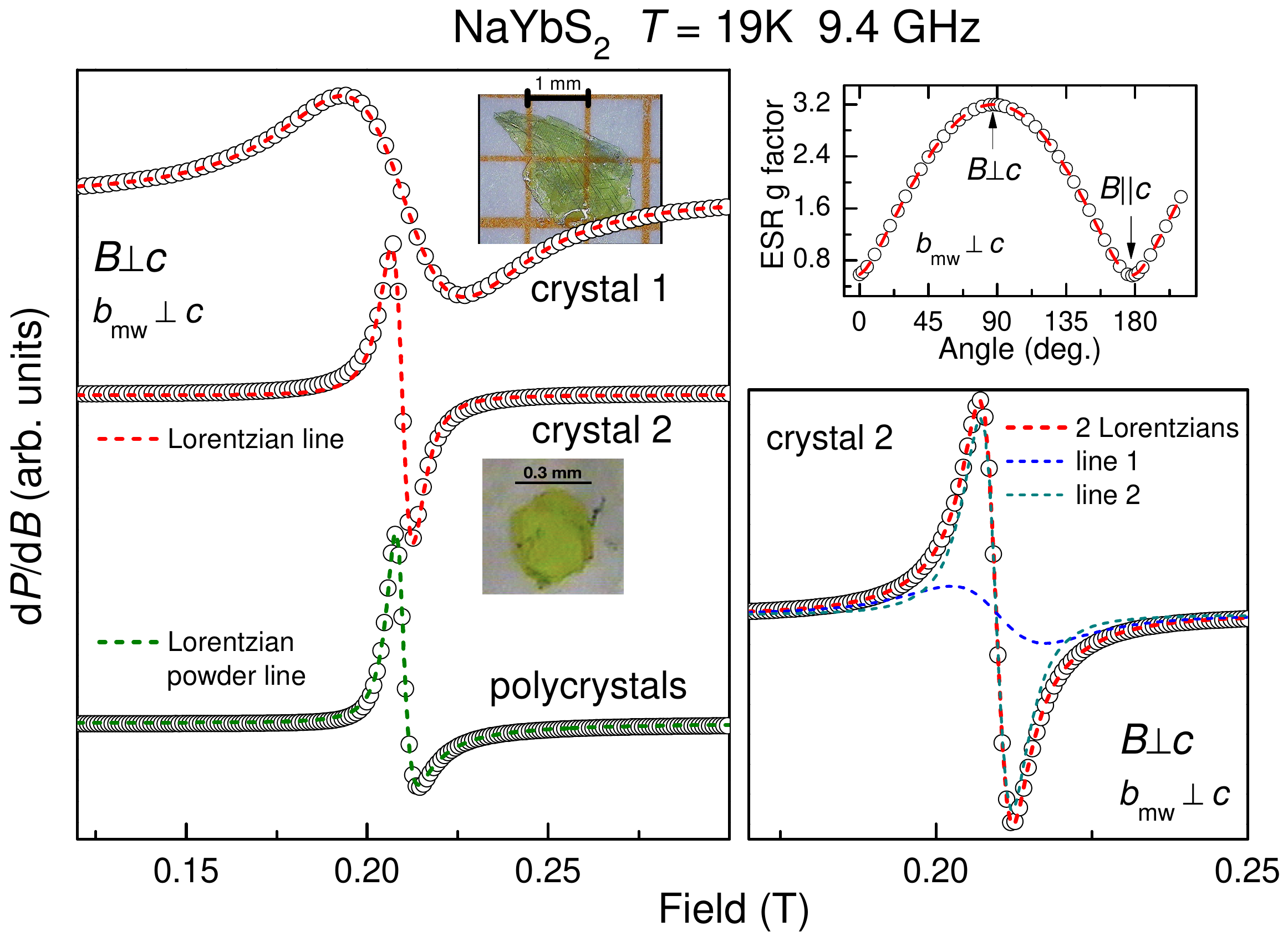}
\end{center}
\caption{
Typical ESR spectra of NaYbS$_{2}$ polycrystals and single crystals (different size, from 2 batches) for the external field $B\bot c$-axis and the microwave field $b_{mw}\bot c$-axis. Dashed lines indicate a Lorentzian lineshape. Lower right frame: the line of the smaller crystal 2 can be well fitted by two Lorentzians at the same resonance field. Upper right frame: Anisotropy of the ESR $g$ factor at $T=19$~K. Dashed line indicates $g(\Theta) = \sqrt{g_\|^2\cos^2\Theta + g_\bot^2\sin^2\Theta}$ with $g_\bot=3.19(5)$ and $g_\|=0.57(3)$. The sample was rotated around an axis lying in the basal plane parallel to $b_{mw}$. 
}
\label{Fig1}
\end{figure}

Typical ESR spectra are shown in Fig. \ref{Fig1}. It is worth to note that all the spectra show very clear amplitudes and relatively narrow linewidths as compared to the ESR spectra of YbMgGaO$_{4}$ \cite{li15a}.
The spectra of the single crystals show a large anisotropy, as determined by the magneto-crystalline anisotropy of Yb$^{3+}$ in a uniaxial crystalline-electric field, similar to what has been observed in various other Yb-systems \cite{kurkin65a,kutuzov08a,kutuzov11a,gruner10a}.
The spectra of the polycrystals do not depend on the orientation of the field - as expected for arbitrarily oriented microcrystallites. In order to rule out any contribution to the ESR from Yb$_{2}$O$_{2}$S we measured this phase separately and found no resonance line.

The narrow ESR spectra of the small crystal 2 can be well fitted with \textit{two} Lorentzian lines. The lower right frame of Figure \ref{Fig1} shows a typical example for this double line fitting which works very well for both $B\bot c$ and $B\|c$ directions. The resonance field of each line is the same which indicates that the Yb$^{3+}$ spins corresponding to each line are locally in the same crystalline electric field environment. The line parameters linewidth $\Delta B$ and intensity $I$ differ by a ratio which is constant at least for temperatures below 30~K: $I_{1}/I_{2}=0.76$, $\Delta B_{1} /\Delta B_{2}=1/3$. That means, assuming all probes having the same effective magnetic moment, that around 57 \% of all spin probes have a larger linewidth.
The reasons for the presence of two lines are not obvious.  Two lattice site positions of Yb ions can be ruled out since they are located at one crystallographic site only (``B site'').  
Any breaking of lattice symmetry leads to additional relaxation channels and therefore the presence of crystalline domains should be taken into account in understanding the two linewidths. 
Another possibility for the presence of two lines might be a buckled arrangement of the YbS$_{6}$ octahedra in the triangular layer. Such buckling may be revealed by low-temperature X-ray data which, however, are not available yet.

\begin{figure}[h]
\begin{center}
\includegraphics[width=0.6\columnwidth]{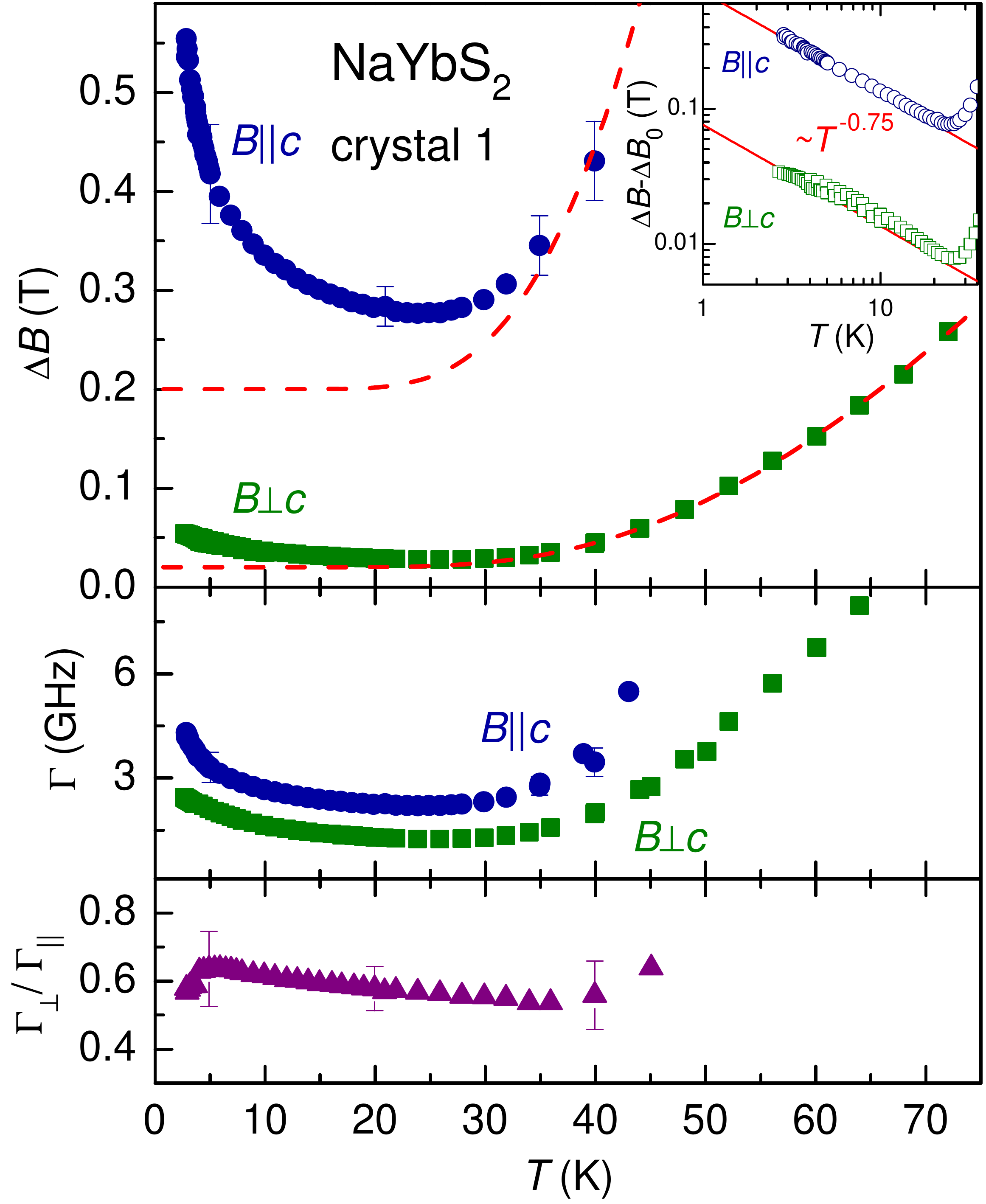}
\end{center}
\caption{
Temperature dependence of ESR linewidth $\Delta B$ in NaYbS$_{2}$ (crystal 1) for two orientations of the external field $B$ to the $c$-axis. Dashed lines indicate  towards higher temperatures the relaxation via the first excited crystalline electric field level of Yb$^{3+}$ at $\Delta=198\pm30$~K. Inset:  Linewidth without a residual contribution ($\Delta B_{0}$ from the dashed lines in the main frame). Solid lines suggest a power law behavior as indicated. Middle frame: relaxation rate $\Gamma=\nu\Delta B/B_{\rm res}$. Lower frame: the ratio of $\Gamma$ along the two directions of the field.
}
\label{Fig2}
\end{figure}

The temperature dependence of the ESR linewidth $\Delta B(T)$ is shown in Fig. \ref{Fig2}. Two regions of temperature behavior can be identified. 
For temperatures above $\approx 30$~K the ESR linewidth broadens due to spin-orbit coupling and the modulation of the ligand field by lattice vibrations. The red dashed lines describe the data with $\Delta B \propto 1/\exp(\Delta/T)-1$ which corresponds to an Orbach process, i.e.~crystalline-electric field splitting of electronic states at an energy $\Delta$ above the ground state are involved in the relaxation \cite{orbach61a,abragam70a}. We obtained $\Delta=198\pm30$~K for all investigated samples. This elevated $\Delta$-value justifies the scenario of an effective spin-1/2 state for the low-temperature regime in agreement with the spin-1/2 entropy of $R\ln2$ found in specific heat results below $\approx30$~K.

The increase of $\Delta B(T)$ towards low temperatures indicates a growing influence of spin correlations. This was as observed, for instance, in detailed ESR measurements of the delafossite PdCrO$_{2}$ for temperatures approaching the magnetic ordering of the Cr$^{3+}$ spins. There, the ESR linewidth could verify a $Z_{2}$-vortex ordering scenario for triangular Heisenberg antiferromagnets \cite{hemmida11a}. For NaYbS$_{2}$, where no magnetic order was observed, a power law increase $\Delta B(T) \propto 1/T^{3/4}$ seems reasonable for all investigated samples as shown in the inset of Fig.~\ref{Fig2}. Such power law behavior indicates a suppression of exchange narrowing by classical critical fluctuations of a 3D order parameter \cite{forster13a}.

In order to characterize the anisotropy of the spin dynamics we consider the relaxation rate $\Gamma$ which is determined by $\Gamma=\nu\Delta B/B_{\rm res}$ (middle and lower frame of Fig. \ref{Fig2}). The rate $\Gamma_{\perp}$ for in-plane field direction is clearly smaller than the out-of-plane rate $\Gamma_{\|}$ - a result which is also found for the $^{11}$Na-NMR relaxation rates for corresponding field directions. The anisotropy $\Gamma_{\perp}/\Gamma_{\|}$ shows no clear temperature dependence down to 2.7~K within experimental error.

\begin{figure}[h]
\begin{center}
\includegraphics[width=0.6\columnwidth]{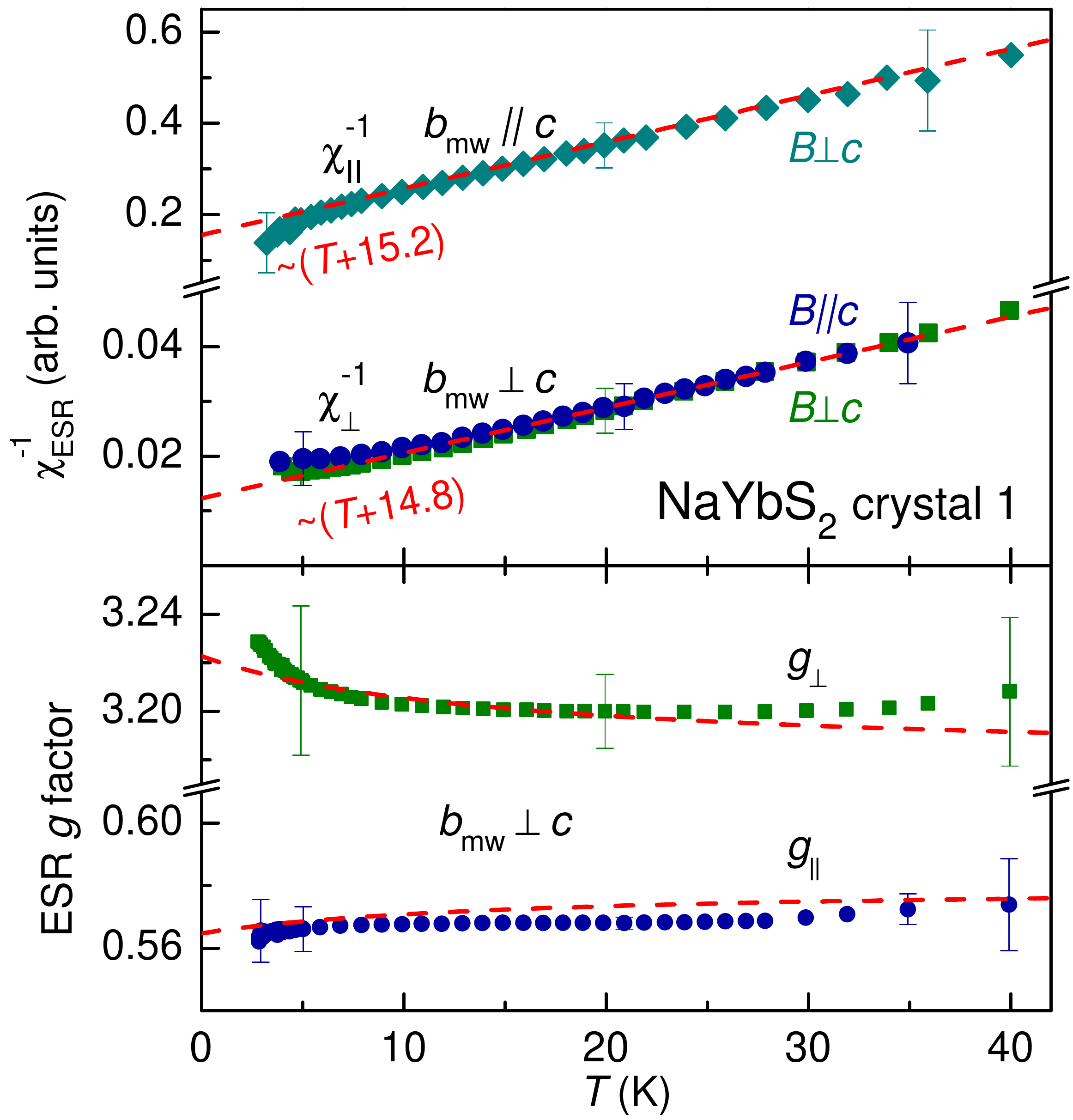}
\end{center}
\caption{
Temperature dependence of ESR intensity $\chi_{\rm ESR}$ and $g$ factor in NaYbS$_{2}$ (crystal 1) for the external field $B$ and the microwave field $b_{mw}$ aligned to the $c$-axis as indicated. 
Dashed lines denote Curie-Weiss fits for the intensity and fits of the $g$-factor according Equations (\ref{eq:1}) and (\ref{eq:2}).
}
\label{Fig3}
\end{figure}

The ESR intensity $I_{\rm ESR}\equiv \chi_{\rm ESR}$ is a direct measure of the spin probe magnetic susceptibility along the direction of the microwave magnetic field $b_{mw}$ \cite{gruner10a}. Figure \ref{Fig3} shows its temperature dependence together with the $g$ factor for in- and out-of-plane directions of the external magnetic field. The dashed lines correspond to a Curie-Weiss behavior  $\chi^{-1}_{\rm ESR}\propto (T+\theta)$. The Weiss temperature $\theta$ is roughly the same for both directions of $b_{mw}$. The indicated values $\theta_\|=15.2$~K and $\theta_\bot=14.8$~K are the values used to fit the temperature dependence of the $g$ factor in the lower frame of Fig. \ref{Fig3}. There the dashed lines refer to a molecular magnetic field description of the anisotropic Yb-Yb interaction \cite{gruner10a,huber09a,sichelschmidt15a} providing a link between the $g$-factor and the exchange anisotropy which is reflected in $\theta_\bot - \theta_\|$ :
\begin{eqnarray}
\label{eq:1}
g_\bot(T) &=& g_{\bot}^0 \cdot \left(1-\frac{\theta_\bot - \theta_\|}{T + \theta_\bot}\right)^\frac{1}{2}\\
\label{eq:2}
g_\|(T) &=& g_{\|}^0 \cdot \left(1+\frac{\theta_\bot - \theta_\|}{T + \theta_\|}\right)
\end{eqnarray}

We obtained $g_{\bot}^0 \approx 3.18$ and $g_{\|}^0 \approx 0.58$ for the adjustable parameters in good agreement with the values describing the anisotropy. Hence, the weak temperature dependence of $g(T)$ demonstrates a small difference $\theta_\bot - \theta_\|$ that cannot be resolved in the $\chi_{\rm ESR}(T)$ data. 

\section{Discussion and Summary}
The presented ESR studies of NaYbS$_{2}$ provide the first results for the local magnetic properties at the Yb$^{3+}$ sites.
The clearly observed uniaxial anisotropy of the $g$ factor (Fig.~\ref{Fig1}) as well as the exponential linewidth increase  (Fig.~\ref{Fig2}) are clear effects of the crystalline electric field that is locally acting on the Yb$^{3+}$ moments. These are centered in edge-sharing, tilted YbS$_{6}$ octahedra 
giving rise to quasi-2D triangular layers of Yb$^{3+}$ ions. For this environment the $J=7/2$ state of the Yb$^{3+}$ ion is characterized by the measured values $g_{\|}$ and $g_{\bot}$ given in Fig.~\ref{Fig3}. 
It is worth to note that the static magnetic susceptibility $\chi$ shows the same uniaxial symmetry as the ESR $g$ factor, see Fig. S8 in \cite{baenitz18a}. However, the anisotropy of $\chi$ is much weaker even if (ESR-silent) Van-Vleck paramagnetic contributions were taken into account \cite{baenitz18a}. 
The small difference $\theta_\bot - \theta_\|$ in the Weiss temperatures obtained from the $g$ factor temperature dependence (Fig.~\ref{Fig3}) is a measure of the anisotropy in the exchange interaction $J_{\mathrm{aniso}}$ between the Yb$^{3+}$ spins. This anisotropy leads to a broadening of the line in contrast to the isotropic exchange $J_{\mathrm{iso}}$ which is responsible for the exchange narrowing mechanism \cite{anderson53a}. With the notation and values as used in Refs. \cite{li15a,baenitz18a} we obtained $J_{\mathrm{iso}}=(4J^{+-}+J^{zz})/3=6.4$~K.  A rough estimate for the anisotropy broadening according to $\Delta B^{\mathrm{aniso}}\propto J_{\mathrm{aniso}}^{2}/J_{\mathrm{iso}}$ with $J_{\mathrm{aniso}}\propto|\theta_\|-\theta_\bot |=0.4$~K yields $\approx 18$~mT. This value is large compared to the maximal estimations for hyperfine- ($\Delta B_{\rm h}=0.7$~mT) and dipolar broadening ($\Delta B_{\rm d}=0.03$~mT) \cite{li15a} using the nearest-neighbour Yb-distance of NaYbS$_{2}$ \cite{baenitz18a}. Hence, in NaYbS$_{2}$ the observed linewidth, reaching the smallest value of 28~mT for $B\bot c$ at 20~K, is largely due to a broadening from anisotropic exchange interactions. Theoretical estimations of the anisotropic exchange parameters from linewidth data should be possible by taking the bonding geometry between neighbouring Yb spins into account. This was done for the bond-dependent exchange in LiCuVO$_{4}$, for instance \cite{krug-von-nidda02a}, and will be subject of detailed future investigations.   

As a summary the static and dynamic magnetic properties of NaYbS$_{2}$ can be nicely investigated by probing locally the Yb$^{3+}$ magnetic moments by ESR. The g values, local susceptibility and the magnetic anisotropy could be determined with an accuracy much higher than so far reported for other spin-liquid candidate compounds \cite{zorko08a,kermarrec14a,padmalekha15a,li15a}.
The observed behavior of the linewidth down to temperatures of 3~K (inset of  Fig.~\ref{Fig2}) as well as the temperature independence of the anisotropy in the spin dynamics (Fig.~\ref{Fig2}) should provide important informations for a microscopic picture of the putative quantum spin-liquid state in NaYbS$_{2}$ \cite{baenitz18a}.

\section*{Acknowledgements} We acknowledge valuable discussions with H. Yasuoka and H.-A. Krug von Nidda. T.D. thanks the Deutsche Forschungsgemeinschaft for financial support in the framework of the CRC 1143.

\section*{References}
\bibliography{JoergBib}

\providecommand{\newblock}{}
\begin{thebibliography}{10}
\expandafter\ifx\csname url\endcsname\relax
  \def\url#1{{\tt #1}}\fi
\expandafter\ifx\csname urlprefix\endcsname\relax\def\urlprefix{URL }\fi
\providecommand{\eprint}[2][]{\url{#2}}

\bibitem{anderson73a}
Anderson P~W 1973 {\em Mat. Res. Bull.\/} {\bf 8} 153

\bibitem{balents10a}
Balents L 2010 {\em Nature\/} {\bf 464} 199

\bibitem{savary17a}
Savary L and Balents L 2017 {\em Rep. Prog. Phys.\/} {\bf 80} 016502

\bibitem{shimizu03a}
Shimizu Y, Miyagawa K, Kanoda K, Maesato M and Saito G 2003 {\em Phys. Rev.
  Lett.\/} {\bf 91}(10) 107001

\bibitem{yamashita08a}
Yamashita S, Nakazawa Y, Oguni M, Oshima Y, Nojiri H, Shimizu Y, Miyagawa K and
  Kanoda K 2008 {\em Nature Physics\/} {\bf 4} 459

\bibitem{yamashita10a}
Yamashita M, Nakata N, Senshu Y, Nagata M, Yamamoto H~M, Kato R, Shibauchi T
  and Matsuda Y 2010 {\em Science\/} {\bf 328} 1246

\bibitem{helton07a}
Helton J~S, Matan K, Shores M~P, Nytko E~A, Bartlett B~M, Yoshida Y, Takano Y,
  Suslov A, Qiu Y, Chung J~H, Nocera D~G and Lee Y~S 2007 {\em Phys. Rev.
  Lett.\/} {\bf 98} 107204

\bibitem{okamoto07a}
Okamoto Y, Nohara M, Aruga-Katori H and Takagi H 2007 {\em Phys. Rev. Lett.\/}
  {\bf 99}(13) 137207

\bibitem{luo18a}
Luo Z~X, Lake E, Mei J~W and Starykh O~A 2018 {\em Phys. Rev. Lett.\/} {\bf
  120} 037204

\bibitem{zorko08a}
Zorko A, Nellutla S, van Tol J, Brunel L~C, Bert F, Duc F, Trombe J~C, de~Vries
  M~A, Harrison A and Mendels P 2008 {\em Phys. Rev. Lett.\/} {\bf 101} 026405

\bibitem{kermarrec14a}
Kermarrec E, Zorko A, Bert F, Colman R~H, Koteswararao B, Bouquet F, Bonville
  P, Hillier A, Amato A, van Tol J, Ozarowski A, Wills A~S and Mendels P 2014
  {\em Phys. Rev. B\/} {\bf 90} 205103

\bibitem{padmalekha15a}
Padmalekha K, Blankenhorn M, Ivek T, Bogani L, Schlueter J and Dressel M 2015
  {\em Physica B: Cond. Mat.\/} {\bf 460} 211 -- 213

\bibitem{li15a}
Li Y, Chen G, Tong W, Pi L, Liu J, Yang Z, Wang X and Zhang Q 2015 {\em Phys.
  Rev. Lett.\/} {\bf 115}(16) 167203

\bibitem{li15b}
Li Y, Liao H, Zhang Z, Li S, Jin F, Ling L, Zhang L, Zou Y, Pi L, Yang Z, Wang
  J, Wu Z and Zhang Q 2015 {\em Scientific Reports\/} {\bf 5} 16419

\bibitem{li17b}
Li Y, Adroja D, Bewley R~I, Voneshen D, Tsirlin A~A, Gegenwart P and Zhang Q
  2017 {\em Phys. Rev. Lett.\/} {\bf 118} 107202

\bibitem{baenitz18a}
Baenitz M, Schlender P, Sichelschmidt J, Onykiienko Y~A, Zangeneh Z, Ranjith
  K~M, Sarkar R, Hozoi L, Walker H~C, Orain J~C, Yasuoka H, van~den Brink J,
  Klauss H~H, Inosov D~S and Doert T 2018 {\em arXiv\/} {\bf 1809} 01947; {\em Phys. Rev. B\/} accepted

\bibitem{campbell76a}
Campbell S~J, Herbert I~R, Warwick C~B and Woodgate J~M 1976 {\em J. Phys. E\/}
  {\bf 9} 443

\bibitem{gruner10a}
Gruner T, Wykhoff J, Sichelschmidt J, Krellner C, Geibel C and Steglich F 2010
  {\em J. Phys.: Condens. Matter\/} {\bf 22} 135602

\bibitem{wykhoff07b}
Wykhoff J, Sichelschmidt J, Lapertot G, Knebel G, Flouquet J, Fazlishanov I~I,
  Krug~von Nidda H~A, Krellner C, Geibel C and Steglich F 2007 {\em Science
  Techn. Adv. Mat.\/} {\bf 8} 389

\bibitem{schleid93a}
Schleid T and Lissner F 1993 {\em Eur. J. Solid State Inorg. Chem.\/} {\bf 30}
  829

\bibitem{kurkin65a}
Kurkin I~N and Shekun L~Y 1965 {\em Opt. Spectrosc.\/} {\bf 18} 738

\bibitem{kutuzov08a}
Kutuzov A, Skvortsova A, Belov S, Sichelschmidt J, Wykhoff J, Eremin I,
  Krellner C, Geibel C and Kochelaev B 2008 {\em J. Phys.: Cond. Matter\/} {\bf
  20} 455208

\bibitem{kutuzov11a}
Kutuzov A and Skvortsova A 2011 {\em J. Phys. Conf. Ser.\/} {\bf 324} 012039

\bibitem{orbach61a}
Orbach R 1961 {\em Proc. Roy. Phys. Soc. A\/} {\bf 264} 458--484

\bibitem{abragam70a}
Abragam A and Bleaney B 1970 {\em Electron Paramagnetic Resonance of Transition
  Ions\/} (Oxford: Clarendon Press)

\bibitem{hemmida11a}
Hemmida M, {Krug von Nidda} H~A and Loidl A 2011 {\em J. Phys. Soc. Jpn.\/}
  {\bf 80} 053707

\bibitem{forster13a}
F\"orster T, Garcia F~A, Gruner T, Kaul E~E, Schmidt B, Geibel C and
  Sichelschmidt J 2013 {\em Phys. Rev. B\/} {\bf 87} 180401

\bibitem{huber09a}
Huber D~L 2009 {\em J. Phys.: Cond. Matter\/} {\bf 21} 322203

\bibitem{sichelschmidt15a}
Sichelschmidt J, Gruner T, Jang D, Steppke A, Brando M, Mitsumoto K and Geibel
  C 2015 {\em J. Phys. Conf. Ser.\/} {\bf 592} 012017

\bibitem{anderson53a}
Anderson P~W and Weiss P~R 1953 {\em Rev. Mod. Phys.\/} {\bf 25} 269

\bibitem{krug-von-nidda02a}
Krug~von Nidda H~A, Svistov L~E, Eremin M~V, Eremina R~M, Loidl A, Kataev V,
  Validov A, Prokofiev A and A\ss{}mus W 2002 {\em Phys. Rev. B\/} {\bf 65}
  134445

\end{thebibliography}

\end{document}